\documentclass[%
reprint,
 amsmath,amssymb,
 aps,
]{revtex4-1}
\usepackage{graphicx}
\usepackage{dcolumn}
\usepackage{bm}

\begin{document}

\preprint{APS/123-QED}

\title{
 Properties of edge states in spin-triplet two-band superconductor}%

\author{Yoshiki Imai}%
\email{imai@phy.saitama-u.ac.jp}%
\affiliation{Theoretische Physik, ETH-H\"onggerberg, CH-8093 Z\"urich, Switzerland}
\affiliation{Department of Physics, Saitama University, Saitama, 338-8570, Japan}
\author{Katsunori Wakabayashi}%
\affiliation{International Center for Materials Nanoarchitectonics (WPI-MANA), National Institute for Materials Science (NIMS), Tsukuba 305-0044, Japan}
\author{Manfred Sigrist}
\affiliation{Theoretische Physik, ETH-H\"onggerberg, CH-8093 Z\"urich, Switzerland}

\date{\today}

\begin{abstract}
Motivated by Sr$_2$RuO$_4$ the magnetic properties of edge states in a two-band spin-triplet superconductor with electron- and hole-like Fermi surfaces are investigated assuming chiral $p$-wave pairing symmetry. The two bands correspond to the $\alpha$-$\beta$-bands of Sr$_2$RuO$_4$
and are modeled within a tight-binding model including inter-orbital hybridization and spin-orbit coupling effects. Including superconductivity the quasiparticle spectrum is determined by means of a self-consistent Bogolyubov-de Gennes calculation.
While a full quasiparticle excitation gap appears in the bulk, gapless states form at the edges which produce spontaneous spin and/or charge currents. The spin current is the result of the specific band structure while the charge current originates from the superconducting condensate. Together they induce a small spin polarization at the edge. Furthermore onsite Coulomb repulsion is included to show that the
edge states are unstable against the formation of a Stoner-like spin polarization of the edge states. Through spin-orbit coupling the current- and the correlation-induced magnetism are coupled to the orientation of the chirality of the superconducting condensate. We speculate that this type of phenomenon could yield a compensation of the magnetic fields induced by currents and also explain the negative result in the recent experimental search for chiral edge currents.
\begin{description}
\item[74.70.Pq,]
\end{description}
\end{abstract}

\pacs{Valid PACS appear here}
\maketitle

\section{Introduction}
Among the unconventional superconductors Sr$_2$RuO$_4$ has received special interest for its unique and most intriguing properties  which resemble in some aspects the spin-triplet superfluid $^3$He~\cite{maen94,mack03}. Despite the relatively low transition temperature of $T_{\rm c} \sim 1.5$ K and the 
fragility of the superconducting phase against disorder, a large bulk of experimental data is 
available nowadays which gives strong evidence that a pairing state with so-called chiral $p$-wave symmetry is realized~\cite{ishi98,luke98}. This state breaks time reversal symmetry implying magnetic properties due an intrinsic angular momentum of the Cooper pairs. Internal magnetism of the superconducting phase has indeed been observed by zero-field relaxation of muon spins~\cite{luke98}. As an odd-parity spin-triplet 
pairing state the gap function of the chiral $p$-wave state can be represented in the vector representation
\begin{equation}
 \bm d (\bm k) = \Delta_0 \hat{z}(k_x\pm {\rm i}k_y), 
\end{equation}
 where the $z$-axis orientation of $ \bm d $ indicates an equal-spin pairing state within the $x$-$y$-plane. This is the analog of the A-phase of $^3$He and is two-fold degenerate, i.e. it has positive or negative chirality depending on the sign of the angular momentum $ L_z $, where $ L_z = \pm 1 $ for the orbital dependences $ k_x\pm {\rm i}k_y $ of the gap function. 
 
The chirality of this pairing state is a topological property and generally yields edge states with chiral properties~\cite{volovik97}. These states are so-called Andreev bound states localized near the surface of a sample penetrating on the length scale of the coherence length~\cite{mats99,mats99e}. These currents carry a finite supercurrent whose direction in the $x$-$y$-plane is connected with the sign of $ L_z $. It has long been suggested that such orbital current might be detectable by sensitive local probes~\cite{volovik85,mats99,furu01}. 
So far all attempts to find the magnetic fields induced by these currents, using scanning Hall probes and SQUID microscopes, have led to negative results~\cite{tame03,kirt07}. On the other hand, the presence of surface subgap states has been demonstrated by quasiparticle tunneling spectroscopy~\cite{ying03,laube00}. Thus, it remains an unresolved puzzle that the predicted currents lie below the detectable limit, in contradiction with the theoretical estimates~\cite{ashby09}. 

It has been early realized that superconductivity in Sr$_2$RuO$_4$ is more complex due to multi-orbital band structure~\cite{agter97}. Indeed it has been suggested that due to the different character of the bands incorporated in this material also the properties of the surface states could be influenced~\cite{ragh10}. 
The three bands crossing the Fermi energy derive from the the $4d$-$t_{2g} $-orbitals of the Ru ions. These
orbitals ($d_{yz} , d_{zx}, d_{xy}$), disperse via $ \pi $-hybridization with intermediate O $ 2p $-orbitals giving rise to the characteristic two-dimensional band structures which can be easily derived from the corresponding tight-binding model in the $x$-$y$-plane of the layered crystal structure, invoking very weak dispersion along the $z$-axis. The three resulting bands separate into two groups the $ \alpha $-$ \beta $-bands belonging to the $ d_{yz} $- and $ d_{zx} $-orbitals with negative parity under reflection at $x$-$y$-plane, in contrast to
the $ d_{xy} $-orbital which yields the $ \gamma $-band. Several experimental results indicate that the $ \gamma $-band with its electron-like Fermi surface is dominant for superconductivity~\cite{mack03}, while
the other two bands are responsible for the strong incommensurate magnetic correlation due to their nearly
one-dimensional character~\cite{mazin97,ng00}. Raghu {\it et al.} have pointed out that in terms of topology the
hole-like $ \alpha $- and the electron-like $ \beta $-band compensate each other such that the Chern numbers of the topology introduced by the chiral $p$-wave state adds up to zero~\cite{ragh10}. Based on this finding they
give an argument for why the surface currents would be unobservable, if these two bands were the 
dominantly superconducting ones. 

In the present study we concentrate also on the $ \alpha $-$\beta$-bands. Our aim is to show how superconducting and magnetic properties of these two bands could
combine in a peculiar way to give rise to novel edge states in the superconducting phase. 
It was noticed earlier that even the normal state has unusual properties 
at surfaces with a normal vector in the $x$-$y$-plane. Spin-orbit coupling combined with the inplane hybridization structure of
the $ d_{yz} $-$d_{zx}$-orbitals gives rise to surface spin currents~\cite{waka07}, which are intimately connected with the presence of a strong anomalous Hall effect proposed for Sr$_2$RuO$_4$~\cite{kont07,kont08,kallin11}. 
Interestingly, 
together with the chiral edge current this spin current 
can generate a finite spin magnetization at the edge, which is related to the spin Hall effect. 
This magnetization may be enhanced due to strong magnetic correlations intrinsic to the $ \alpha $-$\beta$-bands. In our study we discuss a simple two-band model featuring the $ \alpha $-$\beta$-bands and incorporating both superconductivity and magnetism.  The $ \gamma $-band may support and induce superconductivity on the $ \alpha $-$\beta$-bands but is ignored in the present treatment for simplicity. The spin magnetism occurring at the edge contributes to the
magnetization at the surface and may, in principle, act as a compensation to the orbital supercurrents to diminish the
signal to be detected by scanning magnetic probes.

This paper is organized as follows. We construct the effective model in the next section. In Sec. \ref{normal}, we show the spin and charge current. In addition to the currents, the band structure and magnetization in the superconducting phase are shown in Sec. \ref{SC-1}. Summary and discussions are given in Sec. \ref{summary}. 

\section{Model}

\begin{figure}[t]
\includegraphics[width=70mm]{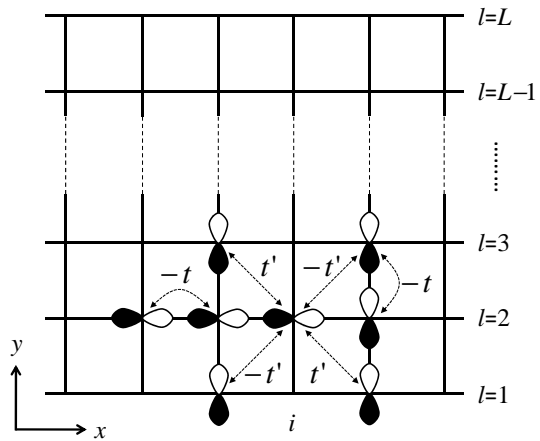}
\caption{Lattice structure of $L$-leg ribbon with two $p_x$ and $p_y$ orbitals. $t$ ($t'$) stands for the hopping integral between same (different) orbitals. }
\label{lattice}
\end{figure}

In this section we introduce the basic model which we investigate for its edge properties. Edges can be most easily addressed in a ribbon-shaped system, an infinite ladder with $ L $ legs as shown in Fig.\ref{lattice}, providing two edges. The two orbitals
can be viewed like $p_x$- and $p_y $-orbitals (Fig.\ref{lattice}) representing $ d_{zx} $- and $d_{yz} $-orbitals, respectively. 
Obviously, the nearest neighbor hopping is only possible via intra-orbital hybridization with a hopping matrix element $ t $. This yields for both orbitals one-dimensional dispersion along the $x$- and $ y$-direction, respectively. 
The inter-orbital hybridization involved next-nearest neighbor hoping of strength (hopping matrix element $ t' $) as depicted
in Fig.\ref{lattice}. 

Including onsite spin-orbit coupling and interactions the corresponding Hamiltonian is written as
\begin{equation}
H = H_{\rm dd}+H_{\rm dd'}+H_{\rm SO}+H_{\rm a}+H_{\rm r}
\label{eqn:ham}
\end{equation}
with
\begin{equation}
\begin{array}{ll}
H_{\rm dd}& \displaystyle =
-t\sum_{i,\sigma}
\left(
 \sum_{l=1}^{L}c^{\dag}_{ilp_x\sigma}c_{i+1lp_x\sigma}\right.  \\
& \displaystyle \hspace{-1mm}+\left.\sum_{l=1}^{L-1}c^{\dag}_{ilp_y\sigma}c_{il+1p_y\sigma}
+ h.c.\right)
-\mu\sum_{ilm\sigma}n_{ilm\sigma}, \\
H_{\rm dd'}& \displaystyle= -t'\sum_{i,m,\sigma}
\left(
 \sum_{l=1}^{L-1}c^{\dag}_{ilm\sigma}c_{i+1l+1\bar{m}\sigma}\right.  \\
&  \displaystyle \hspace{6mm}\left.-\sum_{l=2}^{L}c^{\dag}_{ilm\sigma}c_{i+1l-1\bar{m}\sigma}+ h.c.
\right), \\
H_{\rm SO}&=  \displaystyle -\sum_{i \sigma}\sum_{l}\left(i\lambda\sigma c^{\dag}_{ilp_x\sigma}c_{ilp_y\sigma}+h.c.\right),\\
H_{\rm a}&=  \displaystyle
\frac{U_a}{2}\sum_{ilm\sigma,\sigma'}\left(n_{ilm\sigma}n_{i+1lm\sigma'}+n_{ilm\sigma}n_{il+1m\sigma'}\right), \\
H_{\rm r}&=  \displaystyle U_r\sum_{ilm}n_{ilm\uparrow}n_{ilm\downarrow}
+K_r\sum_{ilm}n_{ilm\uparrow}n_{il\bar{m}\downarrow}  \\
&  \displaystyle +(K_r-J_r)\sum_{ilm(<\bar{m})\sigma}n_{ilm\sigma}n_{il\bar{m}\sigma},
\label{hamiltonian}
\end{array}
\end{equation}
where $c^{\dag}_{ilm\sigma}$ ($c_{ilm\sigma}$) is a creation (annihilation) operator for electron as site $(i,l)$, orbital $m = (p_x, p_y$) and spin $\sigma$ where $ i $ represents the coordinate along the $x$-direction and $ l $ the leg index and $ 
\bar{m} = p_y, p_x )$. Moreover, $\mu$ and $\lambda$ denote the chemical potential and the spin-orbit coupling strength. 
Several interaction terms are introduced. The effective attractive nearest neighbor interaction term 
with $U_a$ ($<0$) will be used
to form $p$-wave Cooper pairs. The repulsive onsite interaction among the electrons includes intra-orbital $U_r$, inter-orbital $K_r$ and Hund's rule coupling $J_r$. For the latter we ignored the exchange and pair hopping part in order to simplify
the discussion without changing the qualitative outcome. We furthermore assume the standard relation $U_r=K_r+2J_r$. 
These repulsive terms are important for the magnetic correlations.
 
\begin{figure}[b]
\includegraphics[width=60mm]{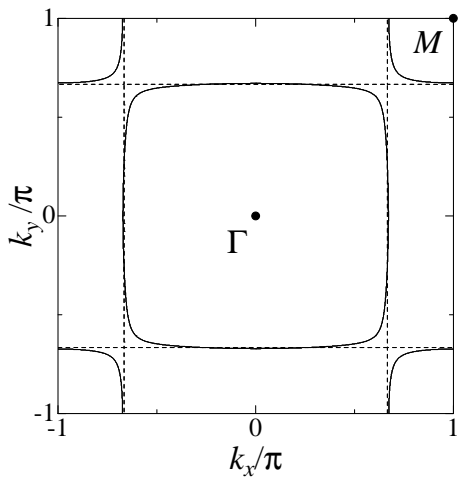}
\caption{Fermi surfaces of two-dimensional bulk system for $t'=0.1t$ and $\lambda=0.1t$ (solid line) and $t'=\lambda=0$ (dashed line), in which particle number $n=8/3$ and $U_a=U_r=J_r=0$.}
\label{fs}
\end{figure}

Neglecting the interaction terms we can determine the band structure leading to the Fermi surfaces shown in Fig.\ref{fs} where the dashed lines indicate the purely one-dimensional Fermi surfaces in the absence of inter-orbital hybridization and spin-orbit coupling ($ t'=\lambda = 0 $). We have adjusted here the chemical potential as to obtain a particle density of $ 8/3 $, close to the experimentally estimated band filling. Obviously even with inter-orbital coupling the nesting property of the Fermi surfaces
prevails, while the electron-like $ \beta $ Fermi surface centered at the $ \Gamma $-point and the hole-like $ \alpha $ Fermi surface centered at the $ M $-point are formed.  

In the following the interaction terms will be decoupled by mean fields. For the attractive nearest-neighbor interaction we
use the BCS-type mean field with the superconducting order parameters defined as
\begin{eqnarray}
\Delta^{x}_{l,m\sigma,m'\sigma'}
  &=&\langle c_{i+1lm\sigma}c_{ilm'\sigma'}\rangle, \\
\Delta^{y}_{l,m\sigma,m'\sigma'}
  &=&\langle c_{il+1m\sigma}c_{ilm'\sigma'}\rangle. 
 \label{eqn:dely}
\end{eqnarray}
Note these mean fields are located on the bonds rather than on the sites. We will restrict here on the spin-triplet channel with inplane equal-spin pairing using only the intra-orbital mean fields $  \Delta^{x}_{l,m\sigma,m\bar{\sigma}} = \delta_{m ,  p_x} \langle c_{i+1lp_x\sigma}c_{ilp_x\bar{\sigma}}\rangle \equiv \Delta^{x}_l$ and $ \Delta^{y}_{l,m\sigma,m \bar{\sigma}} = \delta_{m ,p_y} \langle c_{il+1p_y\sigma}c_{ilp_y\bar{\sigma}}\rangle \equiv \Delta^{y}_l$ while all other pairing decouplings are assumed to vanish. This leads to the chiral $p$-wave phase as the most stable superconducting phase.  
Since the definition of $ \Delta^{y}_l$ is not symmetric with respect to $L/2$, the redefinition of the gap function is given by 
\begin{eqnarray}
\Delta^{y'}_l=
\left\{
\begin{array}{cc}
\Delta^{y}_l/2&(l=1)\\
\Delta^{y}_l/2&(l=L)\\
(\Delta^{y}_{l-1}+\Delta^{y}_l)/2&({\rm otherwise})
\end{array}
\right..
\end{eqnarray}

For the onsite interactions we apply the usual Hartree-Fock approximation with particle density and the spin polarization as follows,
\begin{eqnarray}
n_{lm}&=&n_{lm\uparrow}+n_{lm\downarrow},\\
m_{lm}&=&n_{lm\uparrow}-n_{lm\downarrow}, 
\end{eqnarray}
where 
\begin{equation}
n_{lm\sigma} = \langle n_{ilm\sigma}\rangle = \langle c_{ilm\sigma}^{\dag} c_{ilm\sigma} \rangle .
\end{equation}
Our ribbon geometry allows us to consider the system as homogeneous along the $ x$-direction, but implies the spatial dependence of these mean fields in the transverse direction.

\section{Spin currents in the normal state}
\label{normal}

Before discussing the superconducting phase we first address the surface properties of the normal state for $U_a=0$.
 To illustrate the behavior of the system qualitatively we use mainly the following model parameters:  the ribbon $L=100$, the inter-orbital hopping matrix element $t'=0.1t$, and the electron density $n=8/3$. 

The topology of intra- and inter-orbital hybridization and spin-orbit coupling 
leads to spin currents at the surface. This can be observed on the level of the single-particle part of
the Hamiltonian so that we neglect the interaction terms. We introduce the momentum $ k $ along the
ribbon and define the new electron operators
\begin{equation}
c_{ilm\sigma} = \frac{1}{\sqrt{L_x}} \sum_{k} c_{klm\sigma} e^{-ik x_i}
\end{equation}
with $ x_i $ being the $x$-coordinate of site $ (i,l) $ ($ L_x$: the length of the ribbon assuming periodic boundary 
conditions along $x$-direction). 

The spin dependent current operator along the $x$-axis consists of two parts, 
\begin{equation}
\hat{j}_{l\sigma} = \hat{j}^{(1)}_{l\sigma}+\hat{j}^{(2)}_{l\sigma},
\end{equation}
corresponding to an intra-orbital contribution through nearest-neighbor hopping along the $x$-axis,
\begin{equation}
\hat{j}^{(1)}_{l\sigma} = \sum_{k} (-2t\sin k)
c^{\dag}_{klp_x\sigma}c_{klp_x\sigma},
\end{equation}
and an inter-orbital part with next-nearest neighbor hopping diagonal in the square plaquettes,
\begin{eqnarray}
\hat{j}^{(2)}_{l\sigma}&=&\sum_{k}
(2{\rm i}t'\cos k) (c^{\dag}_{klp_x\sigma}c_{kl+1p_y\sigma}+c^{\dag}_{klp_y\sigma}c_{kl+1p_x\sigma}\nonumber \\
&&\hspace{10mm}-c^{\dag}_{kl+1p_x\sigma}c_{klp_y\sigma}-c^{\dag}_{kl+1p_y\sigma}c_{klp_x\sigma}).
\end{eqnarray}

The expression given here for $\hat{j}^{(2)}_{l\sigma}$ is not symmetric  with respect to $l=L/2$. Therefore we
redefine it as
\begin{eqnarray}
\hat{j}^{(2)'}_{l\sigma}=
\left\{
\begin{array}{cc}
\hat{j}^{(2)}_{l\sigma}/2&(l=1)\\
\hat{j}^{(2)}_{l-1\sigma}/2&(l=L)\\
(\hat{j}^{(2)}_{l-1\sigma}+\hat{j}^{(2)}_{l\sigma})/2&({\rm otherwise})
\end{array}
\right..
\end{eqnarray}
so that  the spin-dependent current operator is now given by
\begin{eqnarray}
\hat{j}_{l\sigma} = \hat{j}^{(1)}_{l\sigma}+\hat{j}^{(2)'}_{l\sigma} ,
\end{eqnarray}
which does not affect the results qualitatively. The charge and spin currents are then given by
\begin{eqnarray}
\hat{J}^{c}_{l\sigma}&=& -\frac{e}{\hbar} \sum_{\sigma} \left\{   \hat{j}^{(1)}_{l\sigma}+\hat{j}^{(2)'}_{l\sigma} \right\}, \\
\hat{J}^{s}_{l\sigma}&=&  \sum_{\sigma} \sigma \left\{   \hat{j}^{(1)}_{l\sigma}+\hat{j}^{(2)'}_{l\sigma} \right\}. 
\end{eqnarray}

\begin{figure}[t]
\includegraphics[width=70mm]{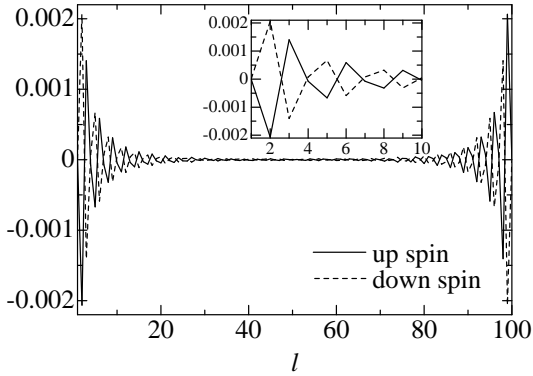}
\caption{Spin-dependent current in normal phase for $L=100$, $t'=\lambda=0.1t$, $U_r=J_r=0$ and $n=8/3$. Solid (dashed) line stands for spin up (down) current. The inset shows magnification near the edge.}
\label{current_normal}
\end{figure}

It is straightforward to determine the groundstate of the normal phase and to calculate the expectation values for the currents. 
The spin-dependent current is depicted in Fig \ref{current_normal}. 
As time reversal symmetry is conserved in the normal phase there is no charge current running. On the other hand, we find a 
spin current at the edges which extends towards to center of the ribbon with a characteristic oscillation. 
\begin{figure}[b]
\includegraphics[width=40mm]{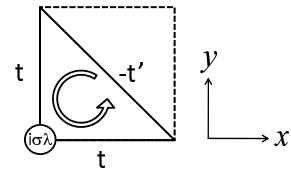}
\caption{Spin-dependent chirality. }
\label{berry}
\end{figure}

The presence of such spin currents can be anticipated by rather simple considerations in the topology of the single-particle Hamiltonian as is revealed, if we examine the phase structure of a square plaquette, as drawn in Fig.\ref{berry}. Separating a single triangle within the plaquette, we see that in the combination of nearest-neighbor intra- and next-nearest neighbor inter-orbital hopping with the onsite spin-orbit coupling  a single particle of spin $ \sigma $ picks up a spin-dependent phase $ \phi_{\sigma} = \sigma \pi / 2 $ moving once around the triangle($\sigma$ being the spin $ \pm 1 $ with quantization axis $z$).
It is straightforward to see that each single particle state carries, in general, a finite spin-dependent currents of equal magnitude around each triangle, but opposite sign for the two spins $\sigma$. This gives rise to a finite spin but a vanishing charge current. 
As they have the same orientation for all four triangles inscribed into the square, the diagonal bonds have no
net current, while the bonds on the square carry a spin current all around. Taking the whole system, these spin currents 
cancel out in the bulk, but remain at the edges due to lack of compensation. 

Alternatively, the phase winding on the triangles can be represented as a ''spin-dependent flux'' $ \Phi_{\sigma} $ density with
$ \Phi_\uparrow = - \Phi_\downarrow $ due to time reversal symmetry.  
We can define a spin dependent current by
\begin{equation}
{\bm j}_{\sigma} ({\bm r}) = {\bm \nabla} \times \hat{z} \Phi_{\sigma} ({\bm r}) 
\end{equation}
which is yields by symmetry $ {\bm j}^c = {\bm j}_\uparrow + {\bm j}_\downarrow = 0 $ and, in general, a non-vanishing spin current density, if
$ \Phi_{\sigma} ({\bm r}) $ is not uniform. The latter condition is satisfied at the edges of our system. 

The oscillations in magnitude of the spin current parallel to the edge can be interpreted as Friedel oscillations as they
correspond roughly to the nesting vector in the band structure, which also is the wave vector of the dominant spin correlation.

\section{Superconducting phase}
\label{SC-1}

In the following we treat the system within the superconducting phase based on the model given by Eqs.(\ref{eqn:ham},\ref{hamiltonian}). For this purpose we will solve the corresponding mean-field in the ribbon with spatial resolution self-consistently for $U_a=-1.5t$ at zero-temperature. 
The other parameters are as in Sec. \ref{normal}. We choose a strong attractive potential $U_a $ in order to obtain a rather small coherence length of a few lattice constants for computational purpose to keep the system size small. 

\subsection{Superconducting order parameter and energy dispersion}
\label{SC-sub-1}

\begin{figure}[t]
\includegraphics[width=70mm]{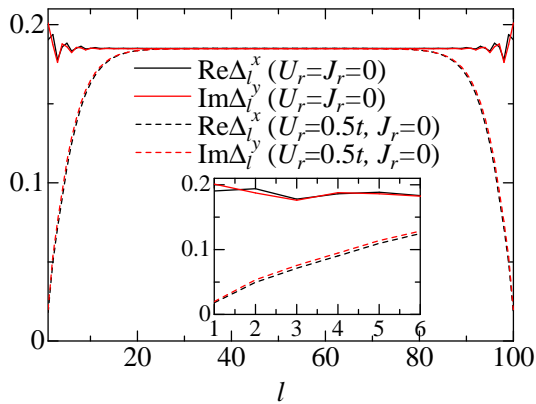}
\caption{(Color online) Order parameters of superconducting state as a function of leg index $l$ for $\lambda =0.1t$ and $U_r=J_r=0$. The black (red) lines stand for $U_r=0.0$ ($U_r=0.5t$). The inset shows magnification near the edge. }
\label{Delta}
\end{figure}

First we analyze the behavior of the superconducting order parameter assuming spin-triplet pairing with inplane equal-spin pairing. In this case the most stable 
pairing state has chiral symmetry in order to avoid nodes in the quasiparticle gap. The order parameter components $ \Delta^x $ and $ \Delta^{y'} $ have a relative
phase $ \pi / 2 $ (or the degenerate $ - \pi/2 $) and are of equal magnitude in the bulk, e.g. Re$\Delta^x$=Im$\Delta^{y'}$, as expected for the chiral $p$-wave state
$\bm d=\hat{z}(k_x+ik_y)$. In the ribbon the order parameter components show a characteristic spatial dependence as depicted in Fig.\ref{Delta} where  Re$\Delta^x $ and Im$\Delta^{y'}$ are shown as a function of the leg index $ l $. The behavior at the edges originates from surface scattering effects, where $ \Delta^x $ (even under reflection at the edge) slightly increases, while $ \Delta^{y'}$ (odd under reflection at the edge) is suppressed~\cite{mats99,furu01}. The pairing state relies on the attractive nearest-neighbor interaction $ H_{\rm a} $, while the repulsive onsite interactions in $ H_{\rm r} $ affects the
order parameter very weakly. Intentionally we keep at this level these onsite interactions weak as they trigger a magnetic instability which we will discuss below.
At this point a few comments are in order. Within our treatment the pairing state with Re$\Delta^x$=$-$Im$\Delta^{y'}$ is degenerate in energy and corresponds to
the time reversed state $\bm d=\hat{z}(k_x-ik_y)$. Spin-orbit interaction does not affect this general fact, as time reversal symmetry is broken spontaneously here. Hereafter we discuss mainly the $\bm d=\hat{z}(k_x+ik_y)$ state unless otherwise noted. 

\begin{figure}[t]
\includegraphics[width=70mm]{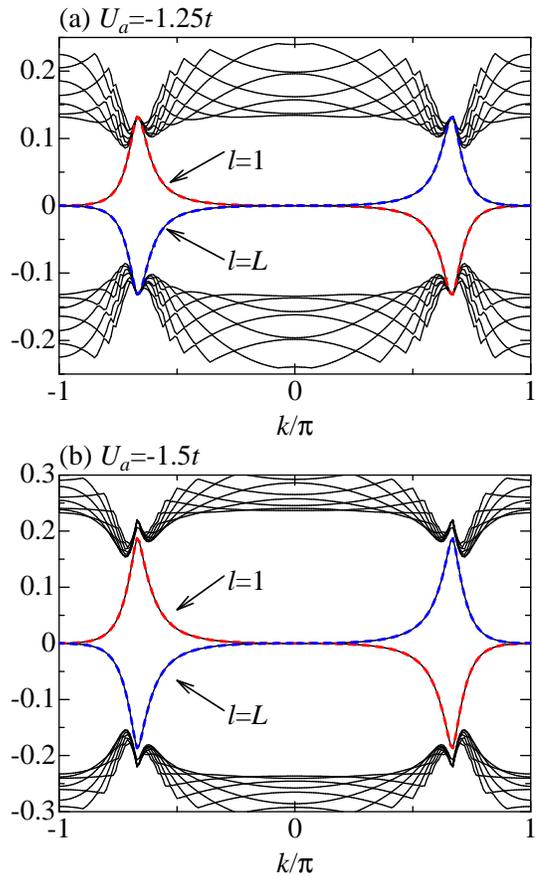}
\caption{(Color online) 
Energy dispersions near low-energy region; (a) $U_a=-1.25t$ and (b) $U_a=-1.5t$ for $U_r=J_r=0$. Red (blue) line stands for edge state from $l=1$ ($l=L$). }
\label{ek}
\end{figure}

Figure \ref{ek} shows the energy spectrum for $U_r=J_r=0$ and $ U_a = -1.25 t $ and $-1.5t$. 
The result shows a gapped ''bulk'' spectrum as expected for the chiral $p$-wave state. In addition two branches of subgap gap states appear corresponding to
Andreev bound states of the two ribbon edges. Especially in Fig. \ref{ek} (b) it can be seen that these edge states are decoupled from the bulk spectrum. 
As the two Fermi surfaces have electron- and hole-like character their topological Chern numbers cancel~\cite{ragh10,imai11}. Therefore each of the two edges 
contributes two zero-energy crossings of opposite chirality, such that these edge states are not topologically protected.

\subsection{Spin and charge currents}
\label{SC-sub-2}

\begin{figure}[t]
\includegraphics[width=70mm]{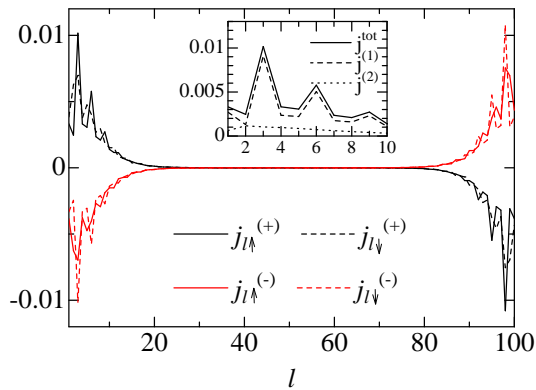}
\caption{(Color online) Spin-dependent currents 
as a function of $l$ for $\lambda =0.1t$ and $U_r=J_r=0$. The black (red) lines stand for $\bm d=\hat{z}(k_x+ik_y)$ ($\bm d=\hat{z}(k_x-ik_y)$) state and the solid (dashed) lines stand for spin up (down) contributions. The inset shows each component of the spin up current for $\bm d=\hat{z}(k_x+ik_y)$ state. }
\label{current}
\end{figure}

We now analyze the equilibrium currents in the absence of onsite repulsion. 
The spin dependent current densities along $x$-directions are shown in Fig.\ref{current} as a function of the leg index $ l $ for both
chiral $p$-wave states $ \bm d=\hat{z}(k_x \pm ik_y)$ as $ j_{l \sigma}^{(\pm)} $. 
Note that since the current from off-diagonal (superconducting) term becomes less than one-tenth of $  j_{l \sigma} $, the contribution is negligible. 

Obviously the currents are associated with the edges. It is important to see that the main contribution to the currents originates from the intra-orbital ($p_x $) component $ j^{(1)}_{l \sigma} $ which exceeds clearly 
the inter-orbital $  j^{(2)}_{l \sigma} $ as shown in the inset of Fig.\ref{current} (for $ \bm d=\hat{z}(k_x + ik_y)$).

Various symmetry properties can be stated for the currents. The relation between positive and negative chirality is reflected by
\begin{equation}
j_{l \sigma}^{(+)} = - j_{l \bar{\sigma}}^{(-)} .
\end{equation}
The sign of the spin-orbit coupling constant $ \lambda $ yields the relation,
\begin{equation}
j_{l \sigma}^{(\pm)} ({\rm sign} \lambda = + 1) = j_{l \bar{\sigma}}^{(\pm)} ({\rm sign} \lambda = - 1) .
\end{equation}

\begin{figure}[b]
\includegraphics[width=70mm]{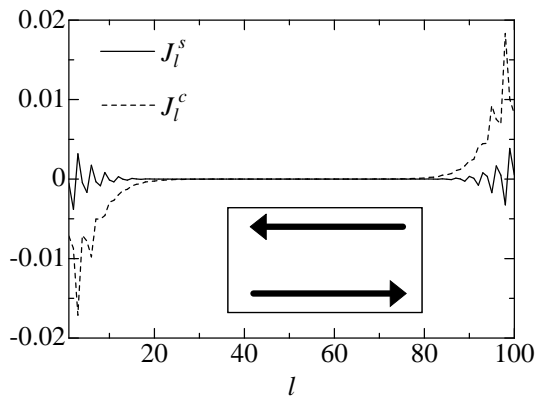}
\caption{Spin and charge currents as a function of $l$ for $\lambda =0.1t$ and $U_r=J_r=0$. The inset shows the flow direction of the chiral charge current near the edges. }
\label{current2}
\end{figure}

We find finite charge and spin current densities which is shown in Fig.\ref{current2}. 
Obviously, the sign of the charge (spin) current depends on the sign of chirality of the superconducting phase (spin-orbit coupling). 

\begin{figure}[t]
\includegraphics[width=70mm]{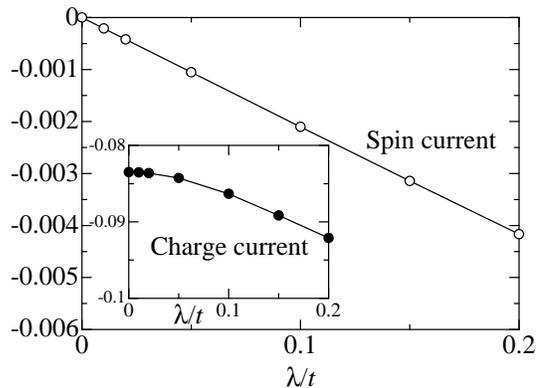}
\caption{Sum of spin current for various $\lambda$. The inset shows the sum of charge current for $U_r=J_r=0$. }
\label{current_sum}
\end{figure}

Figure \ref{current_sum} shows the sums of spin and charge currents up to $l=L/2$, which is defined as 
\begin{eqnarray}
J^{s(c)}_{\rm sum}=\sum_{l=1}^{L/2}J^{s(c)}_{l}.
\end{eqnarray}
When the number of legs is sufficiently large, $J^{s(c)}_{\rm sum}$ is saturated with increasing $l$ from the edges, which corresponds to the total current at each edge. 
In comparison with the result of the charge current, the sum of spin current strongly depends on the amplitude of the spin-orbit interaction.

\begin{figure}[b]
\includegraphics[width=70mm]{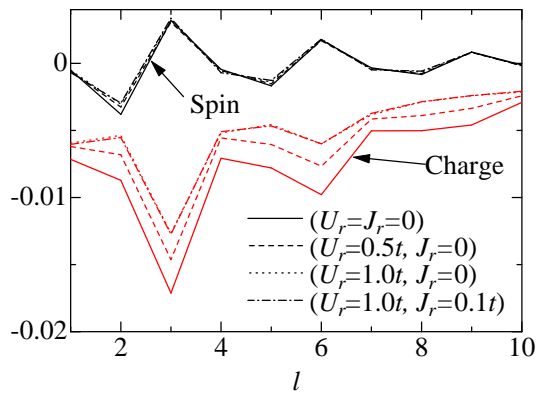}
\caption{(Color online) Spin and charge currents for with/without repulsive interactions near an edge for $\lambda =0.1t$. }
\label{current3}
\end{figure}

Next, the effect of repulsive interactions on the spin and charge currents is discussed. Figure \ref{current3} shows the spin and charge currents with/without repulsive interactions. 
Although the amplitude of each current is slightly modified by repulsive interactions, the sign of the current at each $l$ never changes. In particular, the spin current is almost independent of the amplitude of repulsive interactions, in which the contribution of the horizontal component ($j^{(1)}_{l\sigma}$) is also dominant in the various repulsive interaction region. 
We stress that the currents near the edges flow mainly between $p_x$ orbitals along the edges and the contribution of repulsive interactions to $p_x$ orbitals is rather inconspicuous.

\subsection{Magnetic Properties at edges}
\begin{figure}[t]
\includegraphics[width=70mm]{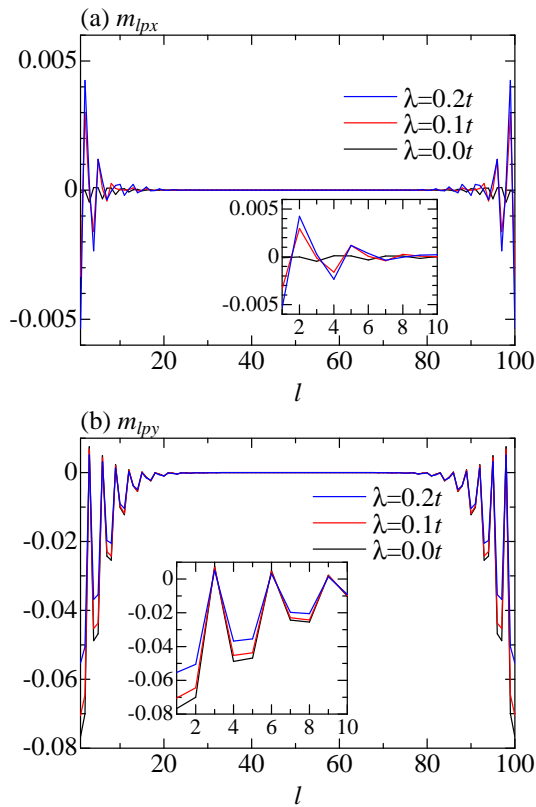}
\caption{(Color online) Spin polarization for various choices of $\lambda$ as a function of $l$. (a) and (b) stand for $m_{lp_x}$ and $m_{lp_y}$, respectively. $U_r=0.5t$ and $J_r=0$.  The insets show the magnifications near edge.}
\label{mag_lambda}
\end{figure}

\begin{figure}[t]
\includegraphics[width=70mm]{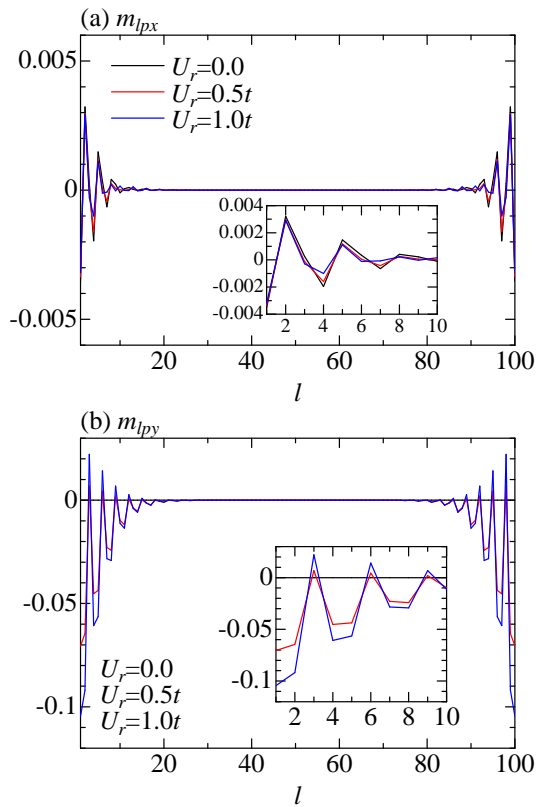}
\caption{(Color online) Spin polarization for several choices of $U_r$. (a) and (b) stand for $m_{lp_x}$ and $m_{lp_y}$, respectively. $\lambda =0.1t$ and $J_r=0$. Both insets show the magnifications near the edge. }
\label{mag_rep}
\end{figure}

The essentially flat electronic bands of the edge states which cover a considerable range of momenta $k$ along the ribbon (see Fig.\ref{ek}) derive mainly from the $ p_y$-orbitals. This special role of the $ p_y $-orbital is also compatible with the suppression of the $ \Delta^{y'} $-component of the superconducting order parameter near the edges which goes hand in hand with the formation of subgap edge states.
These edge states contribute a large density of states at zero energy at the edges and are, therefore, very susceptible to a ''Fermi surface'' instability. Indeed, turning on the repulsive interaction $ U_r $ yields a finite spin magnetization at the edges. The magnetization is uniform along the ribbon and extends towards the interior with an oscillation characteristic to the nesting vector of the band structure (see Fig.\ref{mag_lambda} and \ref{mag_rep}). 

Figure \ref{mag_lambda} shows the spin polarization for various choices of the spin-orbit interaction for  finite repulsive interaction $ U_r $. 
The spin polarization of $p_x$ orbitals for $\lambda=0$ is very small, but increases with growing spin-orbit coupling $ \lambda $ near the edges. On the other hand, the spin-orbit interaction does not affect the  spin polarization $m_{lp_y}$ much. Note, that the two moments $m_{lp_x}$ and $m_{lp_y}$ associated with the two orbitals are connected through both spin-orbit coupling and the next-nearest neighbor hopping $ t'$.  

This magnetic instability for the $p_y$-orbital is of Stoner-like nature, introducing a spin splitting of the surface states due to the exchange interaction. This aspect is clearly visible in Fig.\ref{mag_rep} which shows the spin polarization for different values of $U_r$, showing a stronger magnetization for larger interaction $ U_r$. In sharp contrast, $m_{lp_x}$ is almost independent of $U_r$. 
Consequently we may conclude that the magnetizations carried by the two orbitals are of different origin. 
Through the system geometry and hopping topology the current-induced spin polarization is dominantly associated with the $ p_x $ orbital, while the Stoner-induced spin polarization can, obviously, be attributed to the $ p_y $-orbitals. 

In the absence of spin-orbit coupling the magnetic order of the $ p_y$-orbitals at the edges has a continuous degeneracy due to $SU(2)$-spin rotation. Spin-orbit coupling pins the orientation of the spin polarization to $z$-axis. This is connected with the fact that spin correlations involving the nesting wave vector yields a susceptibility favoring the $z$-axis polarization~\cite{ng00}.
Thus, the remaining two-fold degeneracy (magnetization parallel and antiparallel to $z$-axis) can now be broken by the chirality of the superconducting phase.
The orientation of the spin polarization $m_{lp_x}$ of $ p_x$-orbital is determined by the direction of the chiral edge current, because $m_{lp_x}$ is the result of the combination of spin and chiral edge current. Through 
 spin-orbit coupling, inter-orbital hybridization and Hund's coupling between the two orbital, the spin polarization $m_{lp_x}$ generates a bias for the orientation of $m_{lp_y}$. Thus, the spin magnetization of both orbitals is tied to the chirality of the superconducting phase.

\subsection{Induced magnetic fields}

\begin{figure}[t]
\includegraphics[width=80mm]{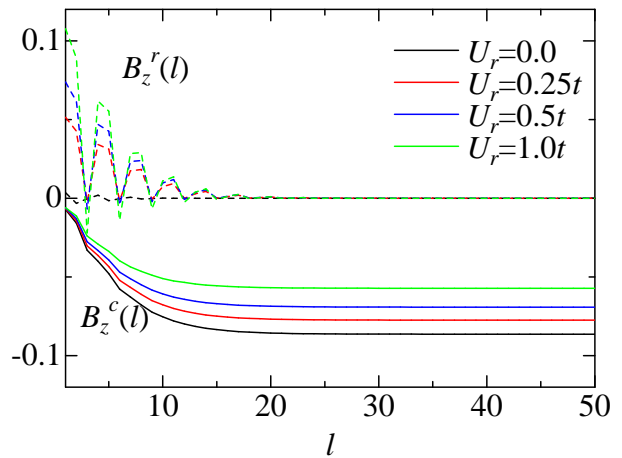}
\caption{(Color online) Sum of spontaneous magnetic field for several choices of repulsive interaction. Solid (dashed) lines stand for the magnetic field from the charge current (spin polarization)  ($\lambda =0.1t$ and $J_r=0$). Note no screening currents have been included in this calculation.}
\label{smf}
\end{figure}

Both the edge charge supercurrent and the spin magnetization generate a net magnetic field. Let us here compare their magnitudes. Using the Maxwell's equation $\nabla \times \bm B= \mu_0 \bm j$ we find that magnetic field of the current is given by
\begin{eqnarray}
B^c_z(l)=\mu_0 \sum_{l'}^{l}\langle J^c_{l'} \rangle=-\frac{et}{\hbar a} \mu_0\sum_{l'}^{l}\langle \tilde{J}^c_{l'}\rangle, 
\end{eqnarray}
where $a$ is the inplane lattice constant, $ \mu_0 $ is the magnetic constant  and $\langle \tilde{J}^c_l \rangle$ is the dimensionless current density along the $x$-direction on layer $l$. The result is shown in Fig.\ref{smf}. Note, that the screening effects have been ignored here such that the magnetic field is constant inside the ribbon. We are only interested in a rough estimate of magnitude of the
field and Meissner screening would introduce counter currents suppressing this field on a length scale of London penetration depth. 

The spin polarization produces a local magnetization to the magnetic field,
\begin{eqnarray}
B^r_z(l)=-\frac{\mu_0}{a^2 c}\mu_{\rm B}(n_{l\uparrow}-n_{l\downarrow}),
\end{eqnarray}
where $\mu_{\rm B}$ is Bohr magneton and $ c$ the lattice constant in $z$-direction. For the mutual comparison of the two fields it is useful to consider the dimensional prefactors. Thus, we consider
\begin{eqnarray}
\frac{\mu_0 et}{\hbar a}\Big/\frac{\mu_0 \mu_{\rm B}}{a^2 c}\sim  \frac{m}{m^*} \frac{c}{a} \sim O(1)
\end{eqnarray}
taking $ \mu_B = \hbar e/2 m $ and $ m^* $ as the effective mass. 

In Fig.\ref{smf} the two magnetic fields can be compared both given in the corresponding units, 
$ \mu_0 et/ \hbar a $ for $ B^c_z $ and $ \mu_0 \mu_B/a^2 c  $ for $B^r_z$. We observed that both
contributions are of similar magnitude and have opposite sign. Since the two units are of similar order
the two fields tend to cancel each other. Note that this compensation occurs for both types of chiral 
domains, since the orientation of the spin magnetization is through spin-orbit coupling coupled with the chirality. 

\section{Summary and discussions}
\label{summary}

Among the three important bands of Sr$_2$RuO$_4$ two, the $ \alpha $- and $ \beta $-bands derived from the $ d_{yz} $ and $ d_{zx}$-orbitals, show remarkable magnetic properties. The special structure of spin-orbit coupling induces here features of an anomalous Hall effect. In particular, the edges carry spin currents. We have analyzed possible consequences of these features on the superconducting phase. Assuming chiral $p$-wave symmetry for the Cooper pairs Andreev bound states at the surface not only induce surface supercurrents, but also introduce a spin polarization. There are two components to the spin magnetism here. One is induced by spin Hall effect like features which give rise to a spin density imbalance at edges when simultaneously charge and spin currents are present. The other component nucleates due to the strong magnetic correlation in the $ \alpha $- and $ \beta $-bands
of incommensurate spin density wave type, which benefit from the modification of quasiparticle spectrum at the surfaces. 

The combination of both shows that the chirality of the superconducting phase and the orientation of the
spin polarizations are coupled. The sign of spin-orbit coupling is so as to lead to opposite magnetizations for the supercurrents and spin magnetic moments. We speculate here that this phenomenon might provide a way to reduce the surface magnetization of the superconducting phase through a compensation effect. In this way the
negative result of the attempts to detect the magnetic fields of a chiral $p$-wave phase by means of
local magnetic probes could be explained. 
One could hope that the observation of the surface spin polarization would be possible by a spin resolved tunneling experiment, which would provide a direct test for our scenario. 

Our model has certainly short-comings as the $ \gamma $-band dominating most likely superconductivity in Sr$_2$RuO$_4$ has been neglected here. Certainly the coupling between the two subsets of bands 
transfers superconductivity from the $ \gamma $- to the $ \alpha $-$\beta$-bands and, 
through spin-orbit and Hund's rule coupling, the spin-polarization is induced in the $ \gamma$-band. 
In any case we expect a reduction of the bare surface current induced local magnetic fields. 
A further issue is the effect of disorder on the discussed effects. Chiral edge states of the $ \alpha $-$\beta$-bands are not topologically protected, since the total Chern number vanishes~\cite{ragh10}. Thus, the situation may be altered by surface roughness and order disorder effects. 
These questions we will be addressed in future studies. We would like to mention also that recently, 
the aspect of the anomalous Hall effect intrinsic to the $ \alpha $-$\beta$-band has recently also been
discussed by Taylor and Kallin~\cite{kallin11} in the context of the polar Kerr effect~\cite{Kapi-Kerr}.

\begin{acknowledgments}
We acknowledge valuable discussions with A. Bouhon, T. Neupert and T. Saso. The work is partly supported by Swiss Nationalfonds and the NCCR MaNEP through the special project "Topomatter". K. W. acknowledges the financial support  of KAKENHI (No. 23310083). 
\end{acknowledgments}

\appendix

\bibliography{paper.bbl}

\providecommand{\noopsort}[1]{}\providecommand{\singleletter}[1]{#1}%
\begin{thebibliography}{24}%
\makeatletter
\providecommand \@ifxundefined [1]{%
 \@ifx{#1\undefined}
}%
\providecommand \@ifnum [1]{%
 \ifnum #1\expandafter \@firstoftwo
 \else \expandafter \@secondoftwo
 \fi
}%
\providecommand \@ifx [1]{%
 \ifx #1\expandafter \@firstoftwo
 \else \expandafter \@secondoftwo
 \fi
}%
\providecommand \natexlab [1]{#1}%
\providecommand \enquote  [1]{``#1''}%
\providecommand \bibnamefont  [1]{#1}%
\providecommand \bibfnamefont [1]{#1}%
\providecommand \citenamefont [1]{#1}%
\providecommand \href@noop [0]{\@secondoftwo}%
\providecommand \href [0]{\begingroup \@sanitize@url \@href}%
\providecommand \@href[1]{\@@startlink{#1}\@@href}%
\providecommand \@@href[1]{\endgroup#1\@@endlink}%
\providecommand \@sanitize@url [0]{\catcode `\\12\catcode `\$12\catcode
  `\&12\catcode `\#12\catcode `\^12\catcode `\_12\catcode `\%12\relax}%
\providecommand \@@startlink[1]{}%
\providecommand \@@endlink[0]{}%
\providecommand \url  [0]{\begingroup\@sanitize@url \@url }%
\providecommand \@url [1]{\endgroup\@href {#1}{\urlprefix }}%
\providecommand \urlprefix  [0]{URL }%
\providecommand \Eprint [0]{\href }%
\providecommand \doibase [0]{http://dx.doi.org/}%
\providecommand \selectlanguage [0]{\@gobble}%
\providecommand \bibinfo  [0]{\@secondoftwo}%
\providecommand \bibfield  [0]{\@secondoftwo}%
\providecommand \translation [1]{[#1]}%
\providecommand \BibitemOpen [0]{}%
\providecommand \bibitemStop [0]{}%
\providecommand \bibitemNoStop [0]{.\EOS\space}%
\providecommand \EOS [0]{\spacefactor3000\relax}%
\providecommand \BibitemShut  [1]{\csname bibitem#1\endcsname}%
\let\auto@bib@innerbib\@empty
\bibitem [{\citenamefont {Maeno}\ \emph {et~al.}(1994)\citenamefont {Maeno},
  \citenamefont {Hashimoto}, \citenamefont {Yoshida}, \citenamefont
  {Nishizaki}, \citenamefont {Fujita}, \citenamefont {Bednorz},\ and\
  \citenamefont {Lichtenberg}}]{maen94}%
  \BibitemOpen
  \bibfield  {author} {\bibinfo {author} {\bibfnamefont {Y.}~\bibnamefont
  {Maeno}}, \bibinfo {author} {\bibfnamefont {H.}~\bibnamefont {Hashimoto}},
  \bibinfo {author} {\bibfnamefont {K.}~\bibnamefont {Yoshida}}, \bibinfo
  {author} {\bibfnamefont {S.}~\bibnamefont {Nishizaki}}, \bibinfo {author}
  {\bibfnamefont {T.}~\bibnamefont {Fujita}}, \bibinfo {author} {\bibfnamefont
  {J.~G.}\ \bibnamefont {Bednorz}}, \ and\ \bibinfo {author} {\bibfnamefont
  {F.}~\bibnamefont {Lichtenberg}},\ }\href@noop {} {\bibfield  {journal}
  {\bibinfo  {journal} {Nature}\ }\textbf {\bibinfo {volume} {372}},\ \bibinfo
  {pages} {532} (\bibinfo {year} {1994})}\BibitemShut {NoStop}%
\bibitem [{\citenamefont {Mackenzie}\ and\ \citenamefont
  {Maeno}(2003)}]{mack03}%
  \BibitemOpen
  \bibfield  {author} {\bibinfo {author} {\bibfnamefont {A.~P.}\ \bibnamefont
  {Mackenzie}}\ and\ \bibinfo {author} {\bibfnamefont {Y.}~\bibnamefont
  {Maeno}},\ }\href@noop {} {\bibfield  {journal} {\bibinfo  {journal} {Rev.
  Mod. Phys.}\ }\textbf {\bibinfo {volume} {75}},\ \bibinfo {pages} {657}
  (\bibinfo {year} {2003})}\BibitemShut {NoStop}%
\bibitem [{\citenamefont {Ishida}\ \emph {et~al.}(1998)\citenamefont {Ishida},
  \citenamefont {Mukuda}, \citenamefont {Kitaoka}, \citenamefont {Asayama},
  \citenamefont {Mao}, \citenamefont {Mori},\ and\ \citenamefont
  {Maeno}}]{ishi98}%
  \BibitemOpen
  \bibfield  {author} {\bibinfo {author} {\bibfnamefont {K.}~\bibnamefont
  {Ishida}}, \bibinfo {author} {\bibfnamefont {H.}~\bibnamefont {Mukuda}},
  \bibinfo {author} {\bibfnamefont {Y.}~\bibnamefont {Kitaoka}}, \bibinfo
  {author} {\bibfnamefont {K.}~\bibnamefont {Asayama}}, \bibinfo {author}
  {\bibfnamefont {Z.~Q.}\ \bibnamefont {Mao}}, \bibinfo {author} {\bibfnamefont
  {Y.}~\bibnamefont {Mori}}, \ and\ \bibinfo {author} {\bibfnamefont
  {Y.}~\bibnamefont {Maeno}},\ }\href@noop {} {\bibfield  {journal} {\bibinfo
  {journal} {Nature (London)}\ }\textbf {\bibinfo {volume} {396}},\ \bibinfo
  {pages} {658} (\bibinfo {year} {1998})}\BibitemShut {NoStop}%
\bibitem [{\citenamefont {Luke}\ \emph {et~al.}(1998)\citenamefont {Luke},
  \citenamefont {Fudamoto}, \citenamefont {Kojima}, \citenamefont {Larkin},
  \citenamefont {Merrin}, \citenamefont {Nachumi}, \citenamefont {Uemura},
  \citenamefont {Maeno}, \citenamefont {Mao}, \citenamefont {Mori},
  \citenamefont {Nakamura},\ and\ \citenamefont {Sigrist}}]{luke98}%
  \BibitemOpen
  \bibfield  {author} {\bibinfo {author} {\bibfnamefont {G.~M.}\ \bibnamefont
  {Luke}}, \bibinfo {author} {\bibfnamefont {Y.}~\bibnamefont {Fudamoto}},
  \bibinfo {author} {\bibfnamefont {K.~M.}\ \bibnamefont {Kojima}}, \bibinfo
  {author} {\bibfnamefont {M.~I.}\ \bibnamefont {Larkin}}, \bibinfo {author}
  {\bibfnamefont {J.}~\bibnamefont {Merrin}}, \bibinfo {author} {\bibfnamefont
  {B.}~\bibnamefont {Nachumi}}, \bibinfo {author} {\bibfnamefont {Y.~J.}\
  \bibnamefont {Uemura}}, \bibinfo {author} {\bibfnamefont {Y.}~\bibnamefont
  {Maeno}}, \bibinfo {author} {\bibfnamefont {Z.~Q.}\ \bibnamefont {Mao}},
  \bibinfo {author} {\bibfnamefont {Y.}~\bibnamefont {Mori}}, \bibinfo {author}
  {\bibfnamefont {H.}~\bibnamefont {Nakamura}}, \ and\ \bibinfo {author}
  {\bibfnamefont {M.}~\bibnamefont {Sigrist}},\ }\href@noop {} {\bibfield
  {journal} {\bibinfo  {journal} {Nature (London)}\ }\textbf {\bibinfo {volume}
  {394}},\ \bibinfo {pages} {558} (\bibinfo {year} {1998})}\BibitemShut
  {NoStop}%
\bibitem [{\citenamefont {Volovik}(1997)}]{volovik97}%
  \BibitemOpen
  \bibfield  {author} {\bibinfo {author} {\bibfnamefont {G.~E.}\ \bibnamefont
  {Volovik}},\ }\href@noop {} {\bibfield  {journal} {\bibinfo  {journal} {JETP
  Lett.}\ }\textbf {\bibinfo {volume} {66}},\ \bibinfo {pages} {522} (\bibinfo
  {year} {1997})}\BibitemShut {NoStop}%
\bibitem [{\citenamefont {Matsumoto}\ and\ \citenamefont
  {Sigrist}(1999{\natexlab{a}})}]{mats99}%
  \BibitemOpen
  \bibfield  {author} {\bibinfo {author} {\bibfnamefont {M.}~\bibnamefont
  {Matsumoto}}\ and\ \bibinfo {author} {\bibfnamefont {M.}~\bibnamefont
  {Sigrist}},\ }\href@noop {} {\bibfield  {journal} {\bibinfo  {journal} {J.
  Phys. Soc. Jpn.}\ }\textbf {\bibinfo {volume} {68}},\ \bibinfo {pages} {994}
  (\bibinfo {year} {1999}{\natexlab{a}})}\BibitemShut {NoStop}%
\bibitem [{\citenamefont {Matsumoto}\ and\ \citenamefont
  {Sigrist}(1999{\natexlab{b}})}]{mats99e}%
  \BibitemOpen
  \bibfield  {author} {\bibinfo {author} {\bibfnamefont {M.}~\bibnamefont
  {Matsumoto}}\ and\ \bibinfo {author} {\bibfnamefont {M.}~\bibnamefont
  {Sigrist}},\ }\href@noop {} {\bibfield  {journal} {\bibinfo  {journal} {J.
  Phys. Soc. Jpn.}\ }\textbf {\bibinfo {volume} {68}},\ \bibinfo {pages} {3120}
  (\bibinfo {year} {1999}{\natexlab{b}})}\BibitemShut {NoStop}%
\bibitem [{\citenamefont {Volovik}\ and\ \citenamefont
  {Gorkov}(1985)}]{volovik85}%
  \BibitemOpen
  \bibfield  {author} {\bibinfo {author} {\bibfnamefont {G.~E.}\ \bibnamefont
  {Volovik}}\ and\ \bibinfo {author} {\bibfnamefont {L.}~\bibnamefont
  {Gorkov}},\ }\href@noop {} {\bibfield  {journal} {\bibinfo  {journal} {Sov.
  Phys. JETP}\ }\textbf {\bibinfo {volume} {61}},\ \bibinfo {pages} {843}
  (\bibinfo {year} {1985})}\BibitemShut {NoStop}%
\bibitem [{\citenamefont {Furusaki}\ \emph {et~al.}(2001)\citenamefont
  {Furusaki}, \citenamefont {Matsumoto},\ and\ \citenamefont
  {Sigrist}}]{furu01}%
  \BibitemOpen
  \bibfield  {author} {\bibinfo {author} {\bibfnamefont {A.}~\bibnamefont
  {Furusaki}}, \bibinfo {author} {\bibfnamefont {M.}~\bibnamefont {Matsumoto}},
  \ and\ \bibinfo {author} {\bibfnamefont {M.}~\bibnamefont {Sigrist}},\
  }\href@noop {} {\bibfield  {journal} {\bibinfo  {journal} {Phys. Rev. B}\
  }\textbf {\bibinfo {volume} {64}},\ \bibinfo {pages} {054514} (\bibinfo
  {year} {2001})}\BibitemShut {NoStop}%
\bibitem [{\citenamefont {Tamegai}\ \emph {et~al.}(2003)\citenamefont
  {Tamegai}, \citenamefont {Yamazaki}, \citenamefont {Tokunaga}, \citenamefont
  {Mao},\ and\ \citenamefont {Maeno}}]{tame03}%
  \BibitemOpen
  \bibfield  {author} {\bibinfo {author} {\bibfnamefont {T.}~\bibnamefont
  {Tamegai}}, \bibinfo {author} {\bibfnamefont {K.}~\bibnamefont {Yamazaki}},
  \bibinfo {author} {\bibfnamefont {M.}~\bibnamefont {Tokunaga}}, \bibinfo
  {author} {\bibfnamefont {Z.}~\bibnamefont {Mao}}, \ and\ \bibinfo {author}
  {\bibfnamefont {Y.}~\bibnamefont {Maeno}},\ }\href@noop {} {\bibfield
  {journal} {\bibinfo  {journal} {Physica C}\ }\textbf {\bibinfo {volume}
  {388-389}},\ \bibinfo {pages} {499} (\bibinfo {year} {2003})}\BibitemShut
  {NoStop}%
\bibitem [{\citenamefont {Kirtley}\ \emph {et~al.}(2007)\citenamefont
  {Kirtley}, \citenamefont {Kallin}, \citenamefont {Hicks}, \citenamefont
  {Kim}, \citenamefont {Liu}, \citenamefont {Moler}, \citenamefont {Maeno},\
  and\ \citenamefont {Nelson}}]{kirt07}%
  \BibitemOpen
  \bibfield  {author} {\bibinfo {author} {\bibfnamefont {J.~R.}\ \bibnamefont
  {Kirtley}}, \bibinfo {author} {\bibfnamefont {C.}~\bibnamefont {Kallin}},
  \bibinfo {author} {\bibfnamefont {C.~W.}\ \bibnamefont {Hicks}}, \bibinfo
  {author} {\bibfnamefont {E.-A.}\ \bibnamefont {Kim}}, \bibinfo {author}
  {\bibfnamefont {Y.}~\bibnamefont {Liu}}, \bibinfo {author} {\bibfnamefont
  {K.~A.}\ \bibnamefont {Moler}}, \bibinfo {author} {\bibfnamefont
  {Y.}~\bibnamefont {Maeno}}, \ and\ \bibinfo {author} {\bibfnamefont {K.~D.}\
  \bibnamefont {Nelson}},\ }\href@noop {} {\bibfield  {journal} {\bibinfo
  {journal} {Phys. Rev. B}\ }\textbf {\bibinfo {volume} {76}},\ \bibinfo
  {pages} {014526} (\bibinfo {year} {2007})}\BibitemShut {NoStop}%
\bibitem [{\citenamefont {Liu}\ \emph {et~al.}(2003)\citenamefont {Liu},
  \citenamefont {Nelson}, \citenamefont {Mao}, \citenamefont {Jin},\ and\
  \citenamefont {Maeno}}]{ying03}%
  \BibitemOpen
  \bibfield  {author} {\bibinfo {author} {\bibfnamefont {Y.}~\bibnamefont
  {Liu}}, \bibinfo {author} {\bibfnamefont {K.}~\bibnamefont {Nelson}},
  \bibinfo {author} {\bibfnamefont {Z.}~\bibnamefont {Mao}}, \bibinfo {author}
  {\bibfnamefont {R.}~\bibnamefont {Jin}}, \ and\ \bibinfo {author}
  {\bibfnamefont {Y.}~\bibnamefont {Maeno}},\ }\href@noop {} {\bibfield
  {journal} {\bibinfo  {journal} {J. Low. Temp. Phys.}\ }\textbf {\bibinfo
  {volume} {131}},\ \bibinfo {pages} {1059} (\bibinfo {year}
  {2003})}\BibitemShut {NoStop}%
\bibitem [{\citenamefont {Laube}\ \emph {et~al.}(2000)\citenamefont {Laube},
  \citenamefont {Goll}, \citenamefont {v.~L{\"o}hneysen}, \citenamefont
  {Fogelstr{\"o}m},\ and\ \citenamefont {Lichtenberg}}]{laube00}%
  \BibitemOpen
  \bibfield  {author} {\bibinfo {author} {\bibfnamefont {F.}~\bibnamefont
  {Laube}}, \bibinfo {author} {\bibfnamefont {G.}~\bibnamefont {Goll}},
  \bibinfo {author} {\bibfnamefont {H.}~\bibnamefont {v.~L{\"o}hneysen}},
  \bibinfo {author} {\bibfnamefont {M.}~\bibnamefont {Fogelstr{\"o}m}}, \ and\
  \bibinfo {author} {\bibfnamefont {F.}~\bibnamefont {Lichtenberg}},\
  }\href@noop {} {\bibfield  {journal} {\bibinfo  {journal} {Phys. Rev. Lett.}\
  }\textbf {\bibinfo {volume} {84}},\ \bibinfo {pages} {1595} (\bibinfo {year}
  {2000})}\BibitemShut {NoStop}%
\bibitem [{\citenamefont {Ashby}\ and\ \citenamefont {Kallin}(2009)}]{ashby09}%
  \BibitemOpen
  \bibfield  {author} {\bibinfo {author} {\bibfnamefont {P.~E.~C.}\
  \bibnamefont {Ashby}}\ and\ \bibinfo {author} {\bibfnamefont
  {C.}~\bibnamefont {Kallin}},\ }\href@noop {} {\bibfield  {journal} {\bibinfo
  {journal} {Phys. Rev. B}\ }\textbf {\bibinfo {volume} {79}},\ \bibinfo
  {pages} {224509} (\bibinfo {year} {2009})}\BibitemShut {NoStop}%
\bibitem [{\citenamefont {Agterberg}\ \emph {et~al.}(1997)\citenamefont
  {Agterberg}, \citenamefont {Rice},\ and\ \citenamefont {Sigrist}}]{agter97}%
  \BibitemOpen
  \bibfield  {author} {\bibinfo {author} {\bibfnamefont {D.~F.}\ \bibnamefont
  {Agterberg}}, \bibinfo {author} {\bibfnamefont {T.~M.}\ \bibnamefont {Rice}},
  \ and\ \bibinfo {author} {\bibfnamefont {M.}~\bibnamefont {Sigrist}},\
  }\href@noop {} {\bibfield  {journal} {\bibinfo  {journal} {Phys. Rev. Lett.}\
  }\textbf {\bibinfo {volume} {78}},\ \bibinfo {pages} {3374} (\bibinfo {year}
  {1997})}\BibitemShut {NoStop}%
\bibitem [{\citenamefont {Raghu}\ \emph {et~al.}(2010)\citenamefont {Raghu},
  \citenamefont {Kapitulnik},\ and\ \citenamefont {Kivelson}}]{ragh10}%
  \BibitemOpen
  \bibfield  {author} {\bibinfo {author} {\bibfnamefont {S.}~\bibnamefont
  {Raghu}}, \bibinfo {author} {\bibfnamefont {A.}~\bibnamefont {Kapitulnik}}, \
  and\ \bibinfo {author} {\bibfnamefont {S.~A.}\ \bibnamefont {Kivelson}},\
  }\href@noop {} {\bibfield  {journal} {\bibinfo  {journal} {Phys. Rev. Lett.}\
  }\textbf {\bibinfo {volume} {105}},\ \bibinfo {pages} {136401} (\bibinfo
  {year} {2010})}\BibitemShut {NoStop}%
\bibitem [{\citenamefont {Mazin}\ and\ \citenamefont {Singh}(1997)}]{mazin97}%
  \BibitemOpen
  \bibfield  {author} {\bibinfo {author} {\bibfnamefont {I.~I.}\ \bibnamefont
  {Mazin}}\ and\ \bibinfo {author} {\bibfnamefont {D.}~\bibnamefont {Singh}},\
  }\href@noop {} {\bibfield  {journal} {\bibinfo  {journal} {Phys. Rev. Lett.}\
  }\textbf {\bibinfo {volume} {79}},\ \bibinfo {pages} {733} (\bibinfo {year}
  {1997})}\BibitemShut {NoStop}%
\bibitem [{\citenamefont {Ng}\ and\ \citenamefont {Sigrist}(2000)}]{ng00}%
  \BibitemOpen
  \bibfield  {author} {\bibinfo {author} {\bibfnamefont {K.}~\bibnamefont
  {Ng}}\ and\ \bibinfo {author} {\bibfnamefont {M.}~\bibnamefont {Sigrist}},\
  }\href@noop {} {\bibfield  {journal} {\bibinfo  {journal} {J. Phys. Soc.
  Jpn.}\ }\textbf {\bibinfo {volume} {69}},\ \bibinfo {pages} {3764} (\bibinfo
  {year} {2000})}\BibitemShut {NoStop}%
\bibitem [{\citenamefont {Wakabayashi}\ and\ \citenamefont
  {Sigrist}(2007)}]{waka07}%
  \BibitemOpen
  \bibfield  {author} {\bibinfo {author} {\bibfnamefont {K.}~\bibnamefont
  {Wakabayashi}}\ and\ \bibinfo {author} {\bibfnamefont {M.}~\bibnamefont
  {Sigrist}},\ }\href@noop {} {\bibfield  {journal} {\bibinfo  {journal} {AIP
  Conf. Proc.}\ }\textbf {\bibinfo {volume} {893}},\ \bibinfo {pages} {1269}
  (\bibinfo {year} {2007})}\BibitemShut {NoStop}%
\bibitem [{\citenamefont {Kontani}\ \emph {et~al.}(2007)\citenamefont
  {Kontani}, \citenamefont {Tanaka},\ and\ \citenamefont {Yamada}}]{kont07}%
  \BibitemOpen
  \bibfield  {author} {\bibinfo {author} {\bibfnamefont {H.}~\bibnamefont
  {Kontani}}, \bibinfo {author} {\bibfnamefont {T.}~\bibnamefont {Tanaka}}, \
  and\ \bibinfo {author} {\bibfnamefont {K.}~\bibnamefont {Yamada}},\
  }\href@noop {} {\bibfield  {journal} {\bibinfo  {journal} {Phys. Rev. B}\
  }\textbf {\bibinfo {volume} {75}},\ \bibinfo {pages} {184416} (\bibinfo
  {year} {2007})}\BibitemShut {NoStop}%
\bibitem [{\citenamefont {Kontani}\ \emph {et~al.}(2008)\citenamefont
  {Kontani}, \citenamefont {Tanaka}, \citenamefont {Hirashima}, \citenamefont
  {Yamada},\ and\ \citenamefont {Inoue}}]{kont08}%
  \BibitemOpen
  \bibfield  {author} {\bibinfo {author} {\bibfnamefont {H.}~\bibnamefont
  {Kontani}}, \bibinfo {author} {\bibfnamefont {T.}~\bibnamefont {Tanaka}},
  \bibinfo {author} {\bibfnamefont {D.~S.}\ \bibnamefont {Hirashima}}, \bibinfo
  {author} {\bibfnamefont {K.}~\bibnamefont {Yamada}}, \ and\ \bibinfo {author}
  {\bibfnamefont {J.}~\bibnamefont {Inoue}},\ }\href@noop {} {\bibfield
  {journal} {\bibinfo  {journal} {Phys. Rev. Lett.}\ }\textbf {\bibinfo
  {volume} {100}},\ \bibinfo {pages} {096601} (\bibinfo {year}
  {2008})}\BibitemShut {NoStop}%
\bibitem [{\citenamefont {Taylor}\ and\ \citenamefont
  {Kallin}(2011)}]{kallin11}%
  \BibitemOpen
  \bibfield  {author} {\bibinfo {author} {\bibfnamefont {E.}~\bibnamefont
  {Taylor}}\ and\ \bibinfo {author} {\bibfnamefont {C.}~\bibnamefont
  {Kallin}},\ }\href@noop {} {\bibfield  {journal} {\bibinfo  {journal}
  {arXiv:1111.4471}\ } (\bibinfo {year} {2011})}\BibitemShut {NoStop}%
\bibitem [{\citenamefont {Imai}\ \emph {et~al.}()\citenamefont {Imai},
  \citenamefont {Wakabayashi},\ and\ \citenamefont {Sigrist}}]{imai11}%
  \BibitemOpen
  \bibfield  {author} {\bibinfo {author} {\bibfnamefont {Y.}~\bibnamefont
  {Imai}}, \bibinfo {author} {\bibfnamefont {K.}~\bibnamefont {Wakabayashi}}, \
  and\ \bibinfo {author} {\bibfnamefont {M.}~\bibnamefont {Sigrist}},\
  }\href@noop {} {\bibinfo  {journal} {in preparation}\ }\BibitemShut {NoStop}%
\bibitem [{\citenamefont {Xia}\ \emph {et~al.}(2006)\citenamefont {Xia},
  \citenamefont {Maeno}, \citenamefont {Beyersdorf}, \citenamefont {Fejer},\
  and\ \citenamefont {Kapitulnik}}]{Kapi-Kerr}%
  \BibitemOpen
\bibfield  {journal} {  }\bibfield  {author} {\bibinfo {author} {\bibfnamefont
  {J.}~\bibnamefont {Xia}}, \bibinfo {author} {\bibfnamefont {Y.}~\bibnamefont
  {Maeno}}, \bibinfo {author} {\bibfnamefont {P.~T.}\ \bibnamefont
  {Beyersdorf}}, \bibinfo {author} {\bibfnamefont {M.~M.}\ \bibnamefont
  {Fejer}}, \ and\ \bibinfo {author} {\bibfnamefont {A.}~\bibnamefont
  {Kapitulnik}},\ }\href@noop {} {\bibfield  {journal} {\bibinfo  {journal}
  {Phys. Rev. Lett.}\ }\textbf {\bibinfo {volume} {97}},\ \bibinfo {pages}
  {167002} (\bibinfo {year} {2006})}\BibitemShut {NoStop}%
\end{thebibliography}%

\end{document}